\documentclass{emulateapj}
\usepackage{epsf}
\usepackage{apjfonts}
\usepackage{xspace}
\usepackage{amsmath}
\usepackage{framed}
\usepackage{color}
\def\HH{{\rm H}}
\def\H2{{{\rm H}_2}}
\def\HI{{\rm H\,I}}
\def\HII{{\rm H\,II}}
\def\HeI{{\rm He\,I}}

\def\CII{{\rm C\,II}}
\def\Fe26{{\rm Fe\,XXVI}}
\def\CVI{{\rm C\,VI}}

\def\F{{\cal F}}
\def\G{\tilde{\cal F}}

\def\dim#1{\mbox{\,#1}}

\def\hide#1{}

\begin{document}

\title{Cooling and Heating Functions of Photoionized Gas}

\author{Nickolay Y.\ Gnedin\altaffilmark{1,2,3} and Nicholas Hollon\altaffilmark{2,3}}  
\altaffiltext{1}{Particle Astrophysics Center, 
Fermi National Accelerator Laboratory, Batavia, IL 60510, USA; gnedin@fnal.gov}
\altaffiltext{2}{Department of Astronomy \& Astrophysics, The
  University of Chicago, Chicago, IL 60637 USA} 
\altaffiltext{3}{Kavli Institute for Cosmological Physics and Enrico
  Fermi Institute, The University of Chicago, Chicago, IL 60637 USA}

\begin{abstract}
Cooling and heating functions of cosmic gas are a crucial ingredient
for any study of gas dynamics and thermodynamics in the interstellar
and intergalactic medium. As such, they have been studied extensively
in the past under the assumption of collisional ionization
equilibrium. However, for a wide range of applications, the local
radiation field introduces a non-negligible, often dominant,
modification to the cooling and heating functions. In the most general
case, these modifications cannot be described in simple terms, and
would require a detailed calculation with a large set of chemical
species using a radiative transfer code (the well-known code Cloudy,
for example). We show, however, that for a sufficiently general
variation in the spectral shape and intensity of the incident
radiation field, the cooling and heating functions can be
\emph{approximated} as depending only on several photoionization rates,
which can be thought of as representative samples of the overall
radiation field. This dependence is easy to tabulate and implement in
cosmological or galactic-scale simulations, thus economically
accounting for an important but rarely-included factor in the
evolution of cosmic gas. We also show a few examples where the
radiation environment has a large effect, the most spectacular of
which is a quasar that suppresses gas cooling in its host halo without
any mechanical or non-radiative thermal feedback.
\end{abstract}


\section{Introduction}
\label{sec:intro}

The ability of cosmic gas to radiate its internal energy (i.e.\
radiative cooling) and to absorb energy from the incident radiation
field (radiative heating) is a primary distinction between the gas and
dark matter; radiative heating and cooling processes are important in
almost every study of gas dynamics or thermodynamics in the
interstellar and intergalactic media. Because of this importance,
cooling processes in the gas have been investigated in numerous prior
studies, appear as central chapters in multiple textbooks, and are
computed by several publicly available codes.

However, while the physics of radiative cooling and heating is well
understood, the actual application of cooling and heating functions
for studies of interstellar and intergalactic gas is surprisingly
incomplete. The classic ``standard cooling function''
\citep[e.g.][]{cf:ct69,cf:rcs76,cf:ss82,cf:gs83,cf:bh89,atom:sd93,cf:ll99,cf:bbc01,cf:ss06,cf:gs07,cf:ssa08}
has indeed been computed and tabulated quite precisely. However, the
``standard cooling function'' is computed under the assumption of pure
collisional ionization equilibrium (CIE), which is not always valid in
the interstellar medium and is \emph{never} valid in the intergalactic
medium \citep[c.f.][]{sims:wss09}. In many astrophysical applications
the incident radiation field introduces significant, often dominant,
modifications to the ``standard cooling function''. On top of that, in
some environments the assumption of the photoionization equilibrium
may not be sufficiently accurate \citep{atom:sd93,cf:ss06}.

\begin{figure}[t]
\epsscale{1.15}
\plotone{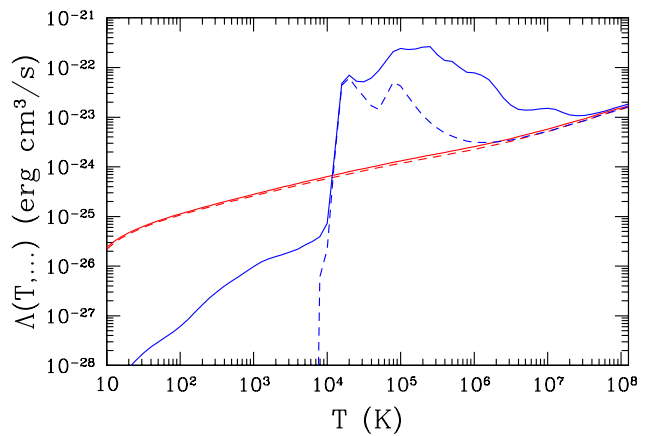}
\caption{Example of the importance of the incident radiation field on
the cooling functions: blue dashed and solid lines show the ``standard
cooling function'' for the metal-free and solar-metallicity gas
respectively. Corresponding red lines show the same cooling functions
for the fully ionized gas.}
\label{fig:ex}
\end{figure}

Such dependence can be illustrated by comparing the pure CIE cooling
function with the cooling function in fully ionized gas, as shown in
Figure \ref{fig:ex} for both metal-free and
solar-metallicity\footnote{Throughout this paper, ``solar
metallicity'' refers to the metallicity of the gas in the solar
neighborhood, $Z\approx 0.02$, not the actual metallicity of the Sun.}
gas. In the fully-ionized limit, where the only cooling process is
bremsstrahlung, the cooling function over a wide range of temperatures
differs from a pure CIE case by \emph{more than two orders of
magnitude}!

Thus, the cooling function for photoionized gas depends not only on
the gas temperature, number density, and metallicity, but also on the
incident radiation field. There is, of course, nothing new in that
statement. The crucial role of the radiation environment has always
been understood by practitioners in the field. The challenge, however,
is in economically accounting for this dependence in full 3D numerical
simulations, where the cooling and heating functions are evaluated
billions or even trillions of times during a single simulation. In the
worst case scenario -- the radiation field $J_\nu$ varies arbitrarily
-- one can introduce sharp edges and features in the radiation
spectrum that are specially designed to ionize particular levels of
particular elements. This allows the cooling function to be
``sculpted'' in an essentially arbitrary way.

One possible way to account for the effect of the incident radiation
field is to fix the radiation spectrum and amplitude. For example, in
studies of intergalactic medium it is often (but not always)
sufficient to account for the cosmic background radiation. Since the
cosmic background radiation evolves with redshift, the cooling and
heating functions become redshift-dependent, but such 1-dimensional
dependence is easy to pre-compute and tabulate for use in simulations
\citep{cf:blbc02,sims:k03,sims:wss09,cf:v11}. Unfortunately, these
cooling and heating rates are then often used for modeling gas
dynamics in galactic halos or even ISM --
environments where the cosmic background radiation is a sub-dominant
component of the incident radiation field.

Therefore, it is desirable to find a way to account for a general
shape of the incident radiation field without the need to recompute
the cooling and heating functions every time they are needed. In this
paper, we show that it is possible to come up with an approximate
solution for this problem using a sufficiently general model for the
radiation field spectrum.

\section{Approximating the Cooling and Heating Functions}
\label{sec:fun}

The radiative term in the internal energy equation - the rate of
change of the gas internal energy due to radiative gains and losses -
can be represented as
\begin{equation}
  \left.\frac{dU}{dt}\right|_{\rm rad} =
  n_b^2\left[\Gamma(T,...) - \Lambda(T,...)\right],
  \label{eq:cf}
\end{equation}
where $U$ is the gas thermal energy and $n_b=n_{\rm H}+4n_{\rm He}+..$
is the total baryon number density. We explicitly factored out $n_b^2$
in both the cooling ($\Lambda$) and the heating ($\Gamma$) functions
so that these are density-independent in the CIE limit.

In the most general case the cooling and heating functions depends on
an extremely large set of arguments: gas temperature $T$, baryon
number density $n_b$ (in addition to $n_b^2$ dependence explicitly
accounted for in Equation (\ref{eq:cf})), the fractional abundance
$X_{ij}$ for the species $i$ (including atomic and ionic species,
various molecules, and cosmic dust) at level $j$, the distribution of
the column density for the species $i$ at level $j$ at different
velocity values with respect to the systemic velocity $dN_{ij}(v)/dv$,
the specific intensity of the radiation field as a function of
frequency $J_\nu$, and the heating rate by cosmic rays $\zeta_{\rm CR}$,
\begin{equation}
  \F(T,...) = \F\left(T,n_b,X_{ij},\frac{dN_{ij}(v)}{dv},J_\nu,\zeta_{\rm
  CR}\right),
  \label{eq:fullcf}
\end{equation}
where hereafter $\F$ denotes either $\Gamma$ or $\Lambda$,
\[
  \F(...) \equiv \left[
    \begin{array}{l}
    \Gamma(...) \phantom{\int}\\
    \Lambda(...)\phantom{\int} 
    \end{array} \right..
\]

Obviously, such a complex dependence cannot be described in simple
terms, and would require a detailed calculation with a large set of
chemical species using a radiative transfer code - for example, the
well-known code Cloudy \citep{misc:fkvf98}. That would make it
impractical as a method for computing the cooling and heating function
in realistic three-dimensional numerical simulations.

We, therefore, adopt several major simplifications. First, we restrict
our focus to a purely optically thin case (all $N_{ij}=0$). Second, we
exclude cooling and heating due to molecules, dust, and cosmic rays, since
these processes crucially depend on radiative transfer and computing
them in the optically thin limit does not make much physical
sense.

With these restrictions, Equation (\ref{eq:fullcf}) becomes
\[
  \F(T,...) = \F(T,n_b,X_{ij},J_\nu).
\]
Even this is way too complex, as the cooling and heating functions
depend on hundreds of individual level populations for atomic and
ionized species. 

In the next simplification step we assume that all atoms and ions are
in the ionization equilibrium, and the level population is in the
equilibrium as well. This assumption is actually valid in a vast
majority of astrophysical environments. In the limit of ionization
equilibrium, the cooling and heating functions only depend on the
total abundance of each chemical element. Finally, if we assume that
the abundance pattern for heavy elements is solar, and ignore small
variations of helium abundance, then the cooling and heating functions
become just functions of the gas metallicity,
\begin{equation}
  \F(T,...) = \F(T,n_b,Z,J_\nu).
  \label{eq:cf4}
\end{equation}

Often, Equation (\ref{eq:cf4}) is what is actually called ``cooling''
and ``heating functions''. For example, the CIE cooling and heating
functions are just
\[
  \F_{\rm CIE}(T,Z) = \F(T,n_b,Z,0)
\]
(which, in this limit, is also independent of $n_b$).

At low enough densities and faint enough incident radiation fields,
most of reactions that result in cooling and heating in gas are
interactions of an atom/ion with either a photon or an
electron. Hence, in this limit cooling and heating functions (Eq.\
\ref{eq:cf}) can be substantially simplified:
\[
  \F(T,n_b,Z,J_\nu) \approx
  \left.\F(T,Z,\frac{J_\nu}{n_b})\right|_{n_b,J_\nu\rightarrow0}.
\]

Unfortunately, this approximation is only valid for rather low values
of the radiation field; for example, it is only marginally valid in
typical ISM conditions in the Milky Way. Several physical processes
break the ideal density-squared dependence. At high enough densities
various 3-body processes become important -- in particular, 3-body
recombination can become important at low temperatures for densities
as low as $10^{-4}\dim{cm}^{-3}$. For hard enough incident radiation
spectra secondary ionizations and heat deposition from secondary
electrons introduce complex density dependence. For strong enough
radiation fields some of highly excited energy levels have critical
densities within the density range we consider here ($n_b \leq
10^6\dim{cm}^{-3}$).

All of these processes, however, are relatively smooth functions of
the density at a constant value of $J_\nu/n_b$. Hence, without
any loss of generality, we can re-write Equation (\ref{eq:cf4}) as
\begin{equation}
  \F(T,...) = \F(T,Z,\frac{J_\nu}{n_b},n_b).
  \label{eq:cffin}
\end{equation}
The advantage of this transformation is that the third argument
includes most of the density dependence; the explicit density
dependence parametrized by the fourth argument is relatively weak and
can be accounted for in a numerical implementation in an economic
manner (the fact we take a full advantage of below).

Finally, we exclude from our cooling and heating functions Compton
cooling and heating and free-free absorption - not because they
violate the ansatz (\ref{eq:cffin}), but because we found empirically
that the functional dependence of those two processes on the
properties of the medium is sufficiently different from the
characteristic dependence of line excitation and emission, so that
numerical approximations that we discuss below become substantially
less accurate with those two processes included. In addition, Compton
processes and free-free absorption depend on the whole shape of the
radiation spectrum, even in the low energy regime that is unimportant
for the line excitation and ionization. These two processes are
described by sufficiently simple analytical expressions that can be
easily added separately to a numerical code if there is such a need.

\section{Modeling Cooling and Heating Functions and the Incident Radiation Field}
\label{sec:rad}

As we mentioned above, it is not possible to account parametrically
for an arbitrary radiation field spectrum. However, in many
astrophysical applications the incident radiation field is dominated
by radiation from stars, AGN or a combination thereof. Thus, we model
the incident radiation field as
\begin{equation}
  J_\nu = J_0 e^{-\tau_\nu} \left[ \frac{1}{1+f_Q}s_\nu +
  \frac{f_Q}{1+f_Q}x^{-\alpha} \right],
  \label{eq:jnu}
\end{equation}
where $x$ is the photon energy in Rydbergs ($x \equiv
h\nu/(1\dim{Ry})$), and $s_\nu$ is a fit to the stellar spectrum from
Starburst99 \citep{misc:lsgd99},
\[
  s_\nu = \frac{1}{5.5}\begin{cases}
  5.5, & x < 1 \\
  x^{-1.8}, & 1 < x < 2.5\\
  0.4x^{-1.8}, & 2.5 < x < 4\\
  2\times10^{-3}x^3/\left(\exp(x/1.4)-1\right), & 4 < x
  \end{cases}  
\]
(this fit is shown in Fig.\ 4 of \citet{ng:rgs02a}). Equation
(\ref{eq:jnu}) also includes the possibility that the incident
radiation field is attenuated by gas with the opacity
\[
  \tau_\nu = \frac{\tau_0}{\sigma_{\HI,0}}\left[
    0.76\sigma_\HI(\nu) + 0.06\sigma_\HeI(\nu) \right],
\]
where $\sigma_{j}(\nu)$ and $\sigma_{j,0}$ are photoionization
cross-sections and their values at respective ionization thresholds
for $j=\HI$ and $\HeI$, and $\tau_0$ is a parameter.

Overall, the radiation field model from Equation (\ref{eq:jnu})
contains 4 parameters: the amplitude $J_0$, the AGN-like power-law
contribution slope $\alpha$, the ratio of the AGN-like to stellar
component $f_Q$, and the shielding optical depth parameter
$\tau_0$. The last 3 parameters are dimensionless; we choose to
measure $J_0$ in units of the typical radiation field in the Milky Way
galaxy, $J_{\rm
MW}=10^6\dim{photons}\,\dim{cm}^{-2}\,\dim{s}^{-1}\,\dim{ster}^{-1}\,\dim{eV}^{-1}$
\citep{ism:d78,ism:mmp83}.

\begin{figure*}[t]
\epsscale{1.15}
\plotone{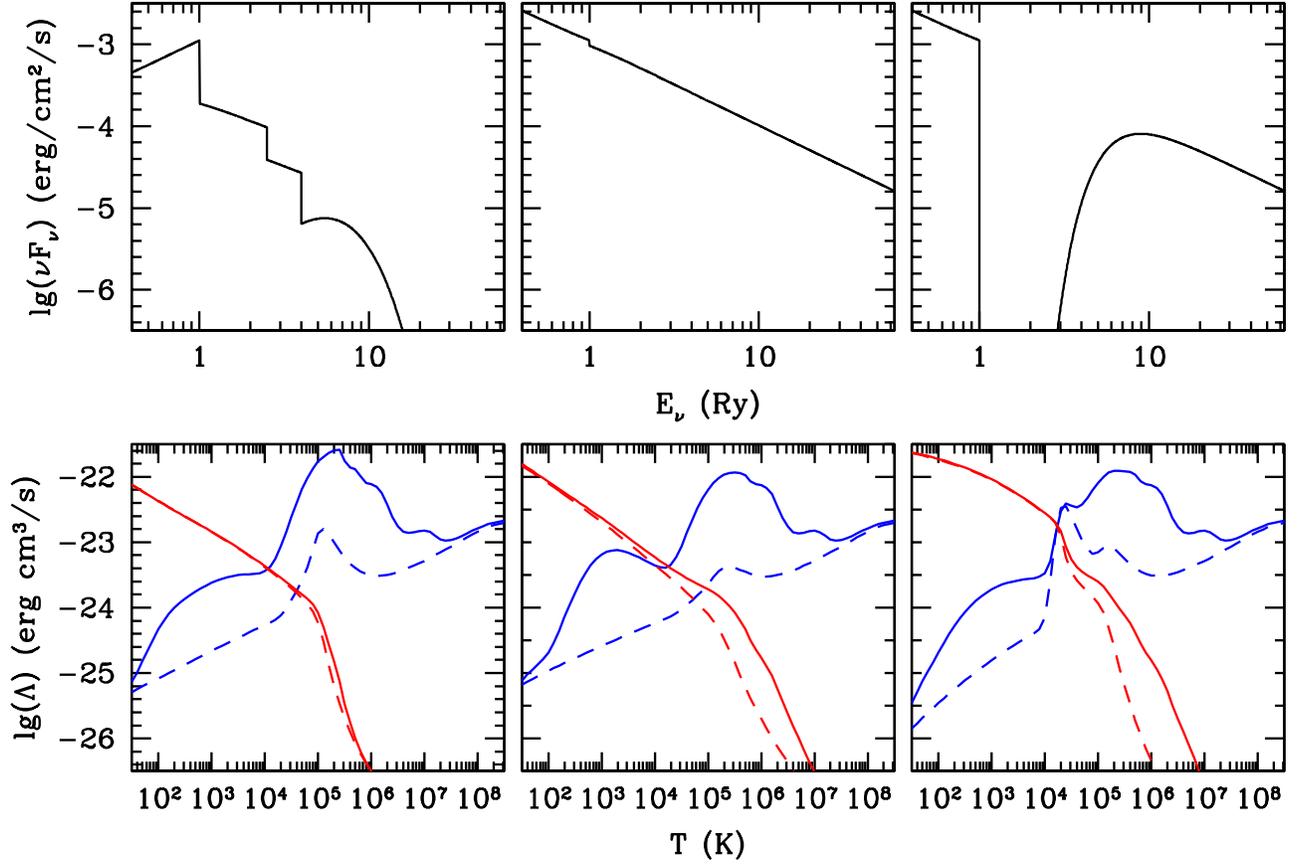}
\caption{Incident radiation fields (top panels) and gas cooling (blue
lines) and heating (red lines) functions (bottom panels) for three
different radiation field models: $[J_0,\alpha,f_Q,\tau_0]=[100J_{\rm
MW},3,10^{-3},0.1]$ (left), $[100J_{\rm MW},2,10,0.1]$ (middle), and
$[100J_{\rm MW},2,10,100]$ (right), all for $n_\HH=1\dim{cm}^{-3}$. As
in Fig.\ \ref{fig:ex}, solid and dashed lines are for $Z=Z_\odot$ and
$Z=0$ respectively. It is clear that the cooling and heating
functions are strongly dependent on the incident radiation field.}
\label{fig:jnu}
\end{figure*}

For each set of parameters, we use the widely known photoionization
code Cloudy \citep{misc:fkvf98} to compute the cooling and heating
function for a range of gas temperatures at fixed gas density and
metallicity. Examples of such computations are shown in Figure
\ref{fig:jnu}. For all 3 cases the radiation field is the same at
$1\dim{Ry}$, but differs in spectral shape at other frequencies (a
stellar spectrum, a power-law spectrum, and a power-law spectrum
shielded by a $\tau_0=100$ cloud). In order to enforce the optically
thin case, we restrict Cloudy calculations to a single zone of a
negligibly small size.

We explore the cooling and heating functions for our radiation field
model by sampling the full parameter space (metallicity, density, and
the radiation field) on the following grid of values:
\begin{eqnarray}
  Z/Z_\odot & = & 0, 0.1, 0.3, 1, 3\nonumber\\
  \lg(n_b/\dim{cm}^{-3}) & = & -6, -5, ..., 6 \nonumber\\
  \lg(J_0\dim{cm}^{-3}/n_b/J_{\rm MW}) & = & -5, -4.5, -4, ..., 7 \nonumber\\ 
  \alpha & = & 0, 0.5, 1, 1.5, 2, 2.5, 3 \label{eq:pars} \\
  \lg(f_q) & = & -3, -2.5, -2, ..., 1 \nonumber\\
  \lg(\tau_0) & = & -1, -0.5, 0, ..., 3 \nonumber
\end{eqnarray}
This parameter range is wide enough to include both extremes shown in
Fig.\ \ref{fig:ex}: the case where the radiation field is completely
negligible and the case where the gas is fully photoionized.

For each of the $5\times13\times25\times7\times9\times9 \approx
920{,}000$ sets of parameters from this grid, we run Cloudy to compute
the cooling and heating functions for 81 values of the temperature
between $10\dim{K}$ and $10^9\dim{K}$ in steps of $0.1\dim{dex}$
(almost 75 million Cloudy runs altogether). Using this large database,
we now consider the various dependencies of the cooling and heating
functions one by one. All of our subsequent approximations are
extensively tested below in \S \ref{sec:test}.

\subsection{Metallicity and Density Dependence}

In a further simplification, we expand both cooling and heating
functions into Taylor series in metallicity up to the quadratic term,
\begin{equation}
  \F \approx \F_0 + \frac{Z}{Z_\odot}\F_1 +
  \left(\frac{Z}{Z_\odot}\right)^2\F_2,
  \label{eq:zexp}
\end{equation}
where all functions $\F_i$ depend only on $T$,
$J_\nu/n_b$, and $n_b$.

We achieve this decomposition in practice by fitting a second degree
polynomial to the five $Z$ values that we sample in Table
(\ref{eq:pars}).
The error introduced by dropping cubic and higher power terms is by
far the smallest of the errors introduced by our approximations -- in
the rms sense, the second order expansion of the Taylor series is
accurate to better than 3\% -- as long as we restrict $Z$ to less than
3 solar metallicities. The quadratic approximation rapidly loses
accuracy as the metallicity increases. At metallicities above
$5Z_\odot$, approximation (\ref{eq:zexp}) even results in negative
cooling functions in a few instances.

Six functions $\Gamma_i$ and $\Lambda_i$ ($i=0,1,2$) can be used
directly, but since cooling and heating functions are not necessarily
monotonic functions of $Z$, some of $\F_1$ and $\F_2$ (again, $\F$
stands for either $\Gamma$ or $\Lambda$) can be negative. Since
interpolation in log-log space is usually more accurate than direct
interpolation, positive functions are much more suitable for
tabulation and interpolation. Hence, we replace 6 functions $\F_i$
with 6 new functions $\G_i$ as
\begin{eqnarray}
  \G_0 & = & \F_0, \nonumber\\
  \G_1 & = & \F_0 + \F_1 + \F_2, \nonumber\\
  \G_2 & = & \F_0 + 2\F_1 + 4\F_2, \nonumber
\end{eqnarray}
where symbol $\G$ also means either the cooling or the heating
function. Functions $\G_i$ are none other than the cooling and heating
functions at $Z=i\times Z_\odot$ and hence are always positive. The
transformation between $\F_i$ and $\G_i$ is linear and can be
trivially inverted.

In the following, we always operate on functions $\G_i$ and convert
them back to $\F_i$ (i.e.\ $\Gamma_i$ and $\Lambda_i$) as the very
last step.

\subsection{Radiation Field Dependence}

So far we still have not resolved the main challenge -- the fact that
the 6 functions $\G_i$ that we need to describe depend on the whole
incident radiation field $J_\nu$,
\[
  \G_i = \G_i(T,\frac{J_\nu}{n_b},n_b).
\]
The primary contribution of this paper is that we further approximate
this dependence by replacing the full radiation field with a finite
set of photoionization rates.

Specifically, let us define a normalized rate $Q_j$ as
\[
  Q_j \equiv \frac{P_j}{n_b},
\]
where $P_j$ is a photoionization rate for some atom or ion,
\[
  P_j = c \int_0^\infty \sigma_j(\nu) n_\nu d\nu,
\]
where $\sigma_j$ is the photoionization cross-section and $n_\nu$ is
the radiation field expressed as the number density of photons at the
frequency $\nu$. We now seek an approximation of the form
\begin{equation}
  \G_i(T,\frac{J_\nu}{n_b},n_b) \approx \G_i(T,Q_j,n_b)
  \label{eq:app}
\end{equation}
for $i=0,1,2$ and some set of $Q_j$. If such an approximation is
possible, then a user would only need to compute the (normalized)
photoionization rates $Q_j$ from his/her assumed radiation field
(provided, such a field is close enough to our assumed shape -  as we
mentioned above, it is always possible to sculpt the cooling and
heating functions by appropriate choosing the radiation field
spectrum). 

We present our specific choice for $Q_j$ in \S \ref{sec:imp}.

\subsection{Explicit Density Dependence}

Finally, we need to address the remaining density dependence in
Equation (\ref{eq:app}). The trick of using $Q_j$ makes this
dependence relatively weak, although highly non-trivial. We adopt the
simplest possible approach -- we tabulate $\G_i$ at the 13 density
values we tested in Table (\ref{eq:pars}) and linearly interpolate in
log-log space. The guaranteed positiveness of $\G_i$ becomes crucial
when working in logarithmic space.

We verified that the error introduced by such interpolation is
completely sub-dominant to the error of our main approximation
(\ref{eq:app}).

\subsection{Notes on the Specific Implementation}
\label{sec:imp}

It makes sense that the rates we choose to represent the radiation
field should sample the wide range of frequencies. For example, since
$\CII$ is an important coolant in the low-temperature regime, one of
the rates should sample the radiation field below the hydrogen
ionization threshold. We choose the photodissociation rate of
molecular hydrogen in the Lyman-Werner band as such a rate, simply
because that rate is also useful for several other processes that can
be modeled in the numerical code (for example, the destruction of
molecular hydrogen). It also makes sense to use the hydrogen and helium
ionization rates since both elements are important coolants at $T \ga
10{,}000\dim{K}$ for all but the highest radiation fields. Finally,
one of the selected rates should be sensitive to high energy photons.

While it is not possible to explore the full set of some 600+
photoionization rates for common chemical elements, we searched the
full hydrogen-like sequence all the way to $\Fe26$ for the most
accurate approximation for the cooling and heating functions.

For a practical numerical implementation, a table with cooling and
heating functions should not exceed about 1 GB of memory, and a much
smaller memory footprint is much preferred. Most modern supercomputers
offer between 1 and 2 GB of memory per computing core; a 1 GB table
would leave no memory for other data structures in a pure MPI code on
a 1 GB/core machine. While pure MPI codes are getting more and more
rare, even a hybrid code that uses just one MPI task per an 8-core
node with 8 GB of memory would encounter difficulty using a table in
excess of 1 GB.

With these constraints in mind, we explored a wide range of possible
3- and 4-dimensional tables. We have not found a 4-dimensional
solution that is sufficiently superior to our best 3-dimensional table
and that fits within our imposed 1 GB memory limit.

Hence, as our primary implementation, we present a 3-dimensional table
that is constructed from four normalized photoionization rates,
$Q_{\rm LW}$, $Q_\HI$, $Q_\HeI$, and $Q_{\rm C\,VI}$, combined into 3
independent parameters,
\begin{eqnarray}
  r_1 & = & Q_{\rm LW}, \nonumber\\
  r_2 & = & \left(\frac{Q_\HI}{Q_{\rm LW}}\right)^{0.353}
            \left(\frac{Q_\HeI}{Q_{\rm LW}}\right)^{0.923}
            \left(\frac{Q_\CVI}{Q_{\rm LW}}\right)^{0.263}, \nonumber\\
  r_3 & = & \left(\frac{Q_\HI}{Q_{\rm LW}}\right)^{-0.103}
            \left(\frac{Q_\HeI}{Q_{\rm LW}}\right)^{-0.375}
            \left(\frac{Q_\CVI}{Q_{\rm LW}}\right)^{0.976}.
  \label{eq:r}
\end{eqnarray}
There exist several other combinations of various rates that result in
almost equivalent approximations. For example, adding $Q_{\rm
Mg\,XII}$ as a fifth rate reduces the error of the approximation by
about 0.03 dex - not significant enough, in our opinion, to justify
computing an extra photoionization rate.

We implement the approximation (\ref{eq:app}) by constructing a grid
of $r_j$ values and computing the average cooling and heating
functions for all incident radiation fields that happen to have the
same values for $r_j$. Specifically, we use a logarithmically-spaced
table for $-14.5 \leq \lg(r_1) \leq -3$, $-9.5 \leq \lg(r_2) \leq
0.5$, and $-8 \leq \lg(r_3) \leq -0.5$ with the logarithmic step of
$0.5\dim{dex}$. We found that using a finer step in the table does not
lead to any increase of accuracy. Such a table includes
$24\times21\times16 = 8064$ entries; each entry contains a sub-grid of
13 density values by 81 temperature values, with 6 numbers $\F_i$ at
each grid point. A full table takes about 192 MB of memory. The table
can be further compressed by eliminating entries with similar cooling
and heating functions, to the total of 3795 entries (92 MB of memory).

With these memory requirements, the final table can be used even in
purely MPI codes on low memory platforms.

\section{Testing the Complete Approximation}
\label{sec:test}

Since we use our sample of Cloudy runs to create the actual tables
with the cooling and heating functions, we need a different data set
to test the accuracy of our approximations. For this purpose we select
100{,}000 points from within our parameter space (\ref{eq:pars}),
sampled uniformly on a logarithmic scale (for the metallicity, we
randomly choose a value between -3 and 0.5 in $\lg(Z)$). For each test
point, we run Cloudy for our 81 values of the temperature to compute
cooling and heating functions. This ``testing'' data set is completely
independent of the data set used to create the tables.

\begin{figure}[t]
\plotone{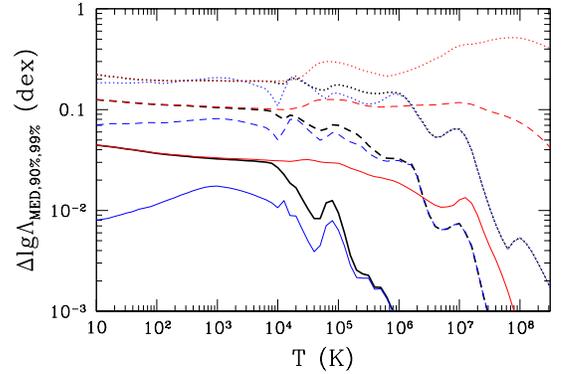}
\caption{Error in approximating the cooling (blue lines), heating (red
lines), and the maximum of the two (black lines) functions with
several photoionization rates. Solid lines trace the median error
(50\% of cases have the error below the solid line), dashed lines
trace the 90\% error, and dotted line trace the 99\% error (only 1\%
of all cases have an error above the dotted line). The existence of
``catastrophic'' errors (a small number of cases with large errors) is
apparent from this figure.}
\label{fig:err}
\end{figure}

We show in Figure \ref{fig:err} the error distribution for our primary
implementation described above. While errors of both functions can be
evaluated, in practice only the difference between the heating and
cooling functions matters (Equation \ref{eq:cf}) - i.e., if one of the
two functions is much larger than the other, a bigger error can be
tolerated for the smaller of the two. Thus, we also show in Figure
\ref{fig:err} the error of the largest of the two functions at each
temperature - that error is a more suitable measure of the actual
error of our approximation.

Two features of Fig.\ \ref{fig:err} are important to note. First, the
median errors for both cooling and heating functions are modest, less
than 10\%. This is very good news indeed, as it shows that the whole
diversity of cooling and heating functions can be parametrized
economically, albeit approximately. Second, unfortunately, is that the
error distribution is not Gaussian, but rather exhibits a long tail
toward large, or ``catastrophic'', errors. For example, in 0.1\% of all
cases that we tested, the error of our approximation reaches a
factor of 2.

\begin{figure}[t]
\plotone{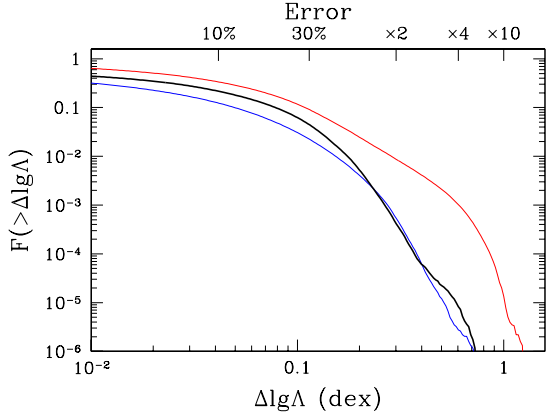}
\caption{Cumulative error distributions for all values of gas
temperatures for the heating (red line), cooling (blue line), and the
maximum of the two (black line) functions for our approximation.}
\label{fig:dist}
\end{figure}

This property of our approximation is further illustrated in Figure
\ref{fig:dist}, where we show the cumulative error distribution for
all temperature values separately for the cooling function, the
heating function, and the maximum of the two (the most appropriate
measure of the actual error). For one case in a million, our
approximation reaches an error of about a factor of 6. Of course, the
specific shapes of the distributions shown in Fig.\ \ref{fig:err} and
\ref{fig:dist} are only applicable to our adopted uniform
sampling. For a specific numerical simulation, the probability of
errors of a particular magnitude will depend on the simulation details
(such as temperature, density, metallicity, and the radiation field
PDFs) and cannot be predicted a priori.

\begin{figure}[t]
\plotone{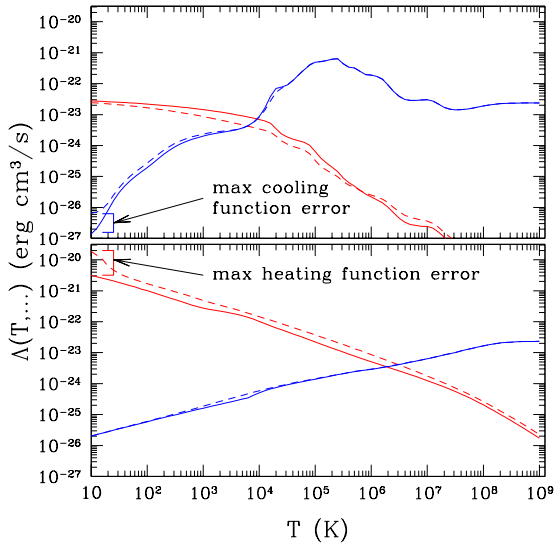}
\caption{Cooling (blue lines) and heating (red lines) functions for
  our test models that maximize the error in the cooling function (top
  panel) and the heating function (bottom panel). Approximate functions
  extracted from our table are shown as dashed lines, while actual
  calculation from Cloudy are shown with solid lines.}
\label{fig:maxerr}
\end{figure}

In order to further illustrate the appearance of a ``catastrophic
error'', we show in Figure \ref{fig:maxerr} the actual cooling and
heating functions that maximize the errors (out of 100{,}000 samples
of 81 temperature values each) in both the heating and cooling
functions. These errors occur in different regions of the parameter
space (i.e. the worst-case heating function is \emph{not} the heating
function that corresponds to the worst-case cooling function, and vice
versa). In most cases that we were able to examine manually, the
largest errors occur at either very low or very high temperatures,
very far from the thermal equilibrium values.

More good news is that the large errors typically occur within a
narrow range of temperatures, like in Fig.\ \ref{fig:maxerr}. Hence,
as soon as the gas temperature changes in the simulation code, the
error in the cooling and heating functions is likely to fall
appreciably. For example, in the bottom panel of Fig.\
\ref{fig:maxerr} the error in the heating function is a factor of 6
for $T=10\dim{K}$. The heating function is very large, so the gas in
this state is going to get heated to higher temperature rapidly,
trying to reach the thermal equilibrium at $T\approx2\times10^6\dim{K}$. As
soon as the gas temperature increases to $T\sim100\dim{K}$, the error in our
approximation drops to less than a factor of 2 (for the fixed values of the
normalized photoionization rates $Q_j$), and will remain within that
range until the thermal equilibrium is reached.

While our adopted spectral shape (\ref{eq:jnu}) is sufficiently
reasonable, it does not represent many situations of astrophysical
interest. In order to test how our approximation fares for other
radiation fields, we apply it to several spectral shapes that are
built-in into Cloudy. Specifically, for each of the built-in spectral
shapes, we randomly choose a value of the density within our full
range $-6\leq\lg(n_b/\dim{cm}^{-3})\leq6$, randomly scale the
radiation field by a factor between $10^{-3}$ and $10^3$ (except for
the Haardt-Madau 2005 background, for which we randomly choose a value
of cosmic redshift between 0 and 6), use Cloudy to compute the actual
cooling and heating functions for that set of parameters, and then
compare Cloudy results with the cooling and heating functions returned
by our approximation when we use the actual photoionization rates that
Cloudy reports for the chosen spectral shape.

\begin{figure}[t]
\plotone{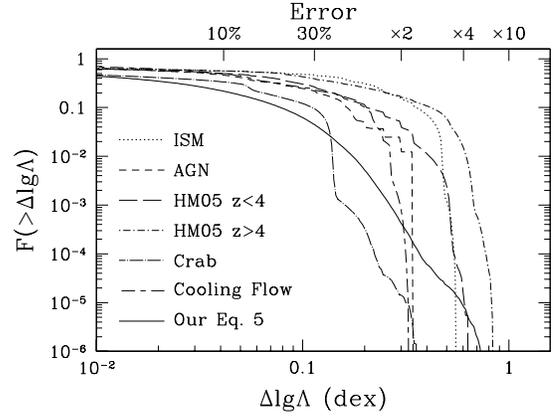}
\caption{Cumulative error distributions for all values of gas
temperatures for the maximum of the cooling and heating functions for
other spectral shapes that are built-in into Cloudy: Milky Way ISM
(Cloudy command ``table ISM'', dotted line), typical AGN spectrum
(Cloudy command ``table AGN'', short-dashed line), Haardt-Madau 2005
cosmic background for $z<4$ (Cloudy command ``table HM05'',
long-dashed line), Haardt-Madau 2005 cosmic background for $z>4$
(Cloudy command ``table HM05'', dot-short-dashed line), continuum from
the Crab nebula (Cloudy command ``table Crab'', dot-long-dashed line),
and continuum from a cooling flow (Cloudy command ``table Cooling
Flow'', short-dash-long-dashed line).  We also show with the solid
line the black line from Fig.\ \ref{fig:dist} for comparison.}
\label{fig:other}
\end{figure}

The cumulative error distribution for several built-in spectral shapes
is shown in Figure \ref{fig:other}. In all cases except one the
maximum error of our approximation does not exceed a factor of 3, and
the error distribution falls off sharply at large errors - in fact,
more sharply than the error distribution for our chosen spectral shape
from Equation (\ref{eq:jnu}); this is not an artifact of incomplete
sampling - we run enough Cloudy models for each spectral shape to
sample the error distribution all the way down to $10^{-6}$.

The only case where our approximation fares worse, producing a
significant fraction ($10^{-3}$) errors as large as a factor of 5, is
the case of the Haardt-Madau 2005 cosmic background for $z>4$. This is
rather disappointing, as cosmological simulations are intended to be
the primary user of our approximation, but we have not succeeded in
finding an approximation that has a noticeably lower error for that
case.

\section{Conclusions}
\label{sec:con}

Our main result is that one can \emph{approximately} represent the
most general cooling and heating functions for gas in ionization
equilibrium as
\begin{equation}
  \left\{\Gamma,\Lambda\right\}(T,n_b,Z,J_\nu) \approx \sum_{i=0}^2
  \left(\frac{Z}{Z_\odot}\right)^i\left\{\Gamma,\Lambda\right\}_i(T,r_j,n_b),
  \label{eq:cfapp}
\end{equation}
where $r_j$ are given in Equation (\ref{eq:r}). This approximation is
rather accurate on average, but suffers from ``catastrophic'' errors
-- in $10^{-6}$ of all cases the approximate cooling or heating
function may deviate from the exact calculation by up to a factor of
6. Thus, our approximation is not suitable for all applications.

Equation (\ref{eq:cfapp}) does capture the qualitative dependence of
the cooling and heating functions on the incident radiation field. To
illustrate this, we show in the appendix three examples where the
cooling and heating functions are significantly modified by the
incident radiation field. The last example -- the quasar irradiating
its own galactic halo (\S \ref{app:agn}) -- not only shows a large
effect the radiation field can have on the cooling/heating functions,
but actually presents an alternative feedback mechanism: the central
black hole suppresses the gas accretion from the halo without any
additional mechanical or thermal feedback.

Our data table and the reader code for it are available online at
{\tt http://astro.uchicago.edu/$\sim$gnedin}

\acknowledgements 

We are grateful to Andrey Kravtsov for enlightening discussions and
constructive criticism. Comments by the anonymous referee helped us to
realize the inadequacy of the original version of our
approximation. This work was supported in part by the DOE at Fermilab,
by the NSF grant AST-0908063, and by the NASA grant NNX-09AJ54G. The
calculations used in this work have been performed on the Joint
Fermilab - KICP Supercomputing Cluster, supported by grants from
Fermilab, Kavli Institute for Cosmological Physics, and the University
of Chicago. We acknowledge the use of code Cloudy \citep{misc:fkvf98}
as the primary research tool of this study. This work made extensive
use of the NASA Astrophysics Data System and {\tt arXiv.org} preprint
server.

\appendix
\section{Some Examples of Cooling and Heating Functions in ISM and IGM}
\label{sec:app}

In this section we present a few examples where the incident radiation
field significantly affects the cooling and heating rates in the
gas. These examples are \emph{not} real physical models, but are
simple demonstrations that the dependence that we explore in this
paper actually matters.

The examples presented here are not exhaustive, of course; one can
imagine many other similar situations. Their purpose is to illustrate
the numerous possible feedback effects in interstellar and
intergalactic environments that arise when we take into account the
effects of the gas metallicity and the incident radiation field on the
cooling and heating rates. These effects can be studied, even if only
approximately, with the approximations presented in this paper.

\subsection{Galactic Halo Near a Quasar}
\label{app:qso}

\begin{figure}[h]
\plottwo{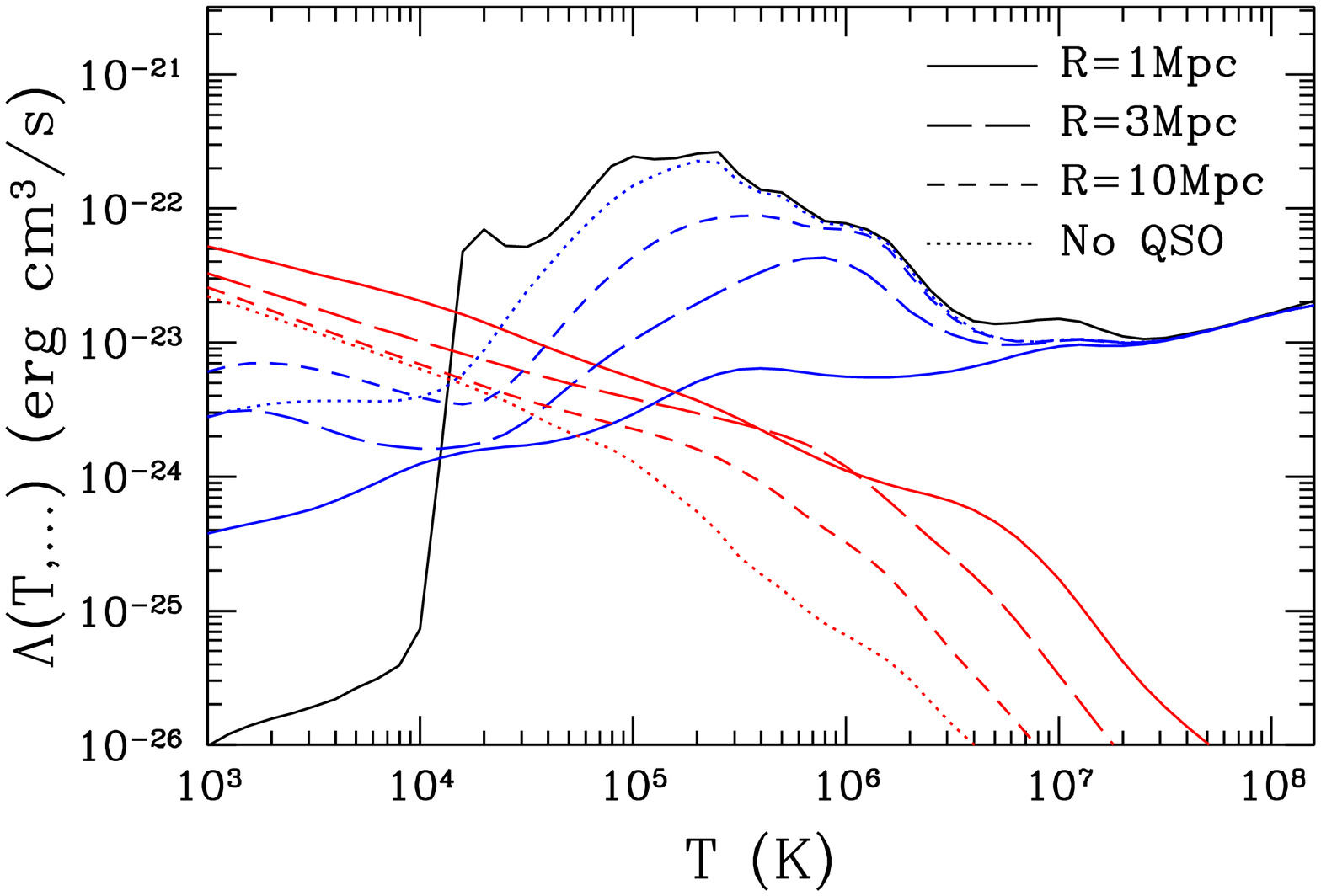}{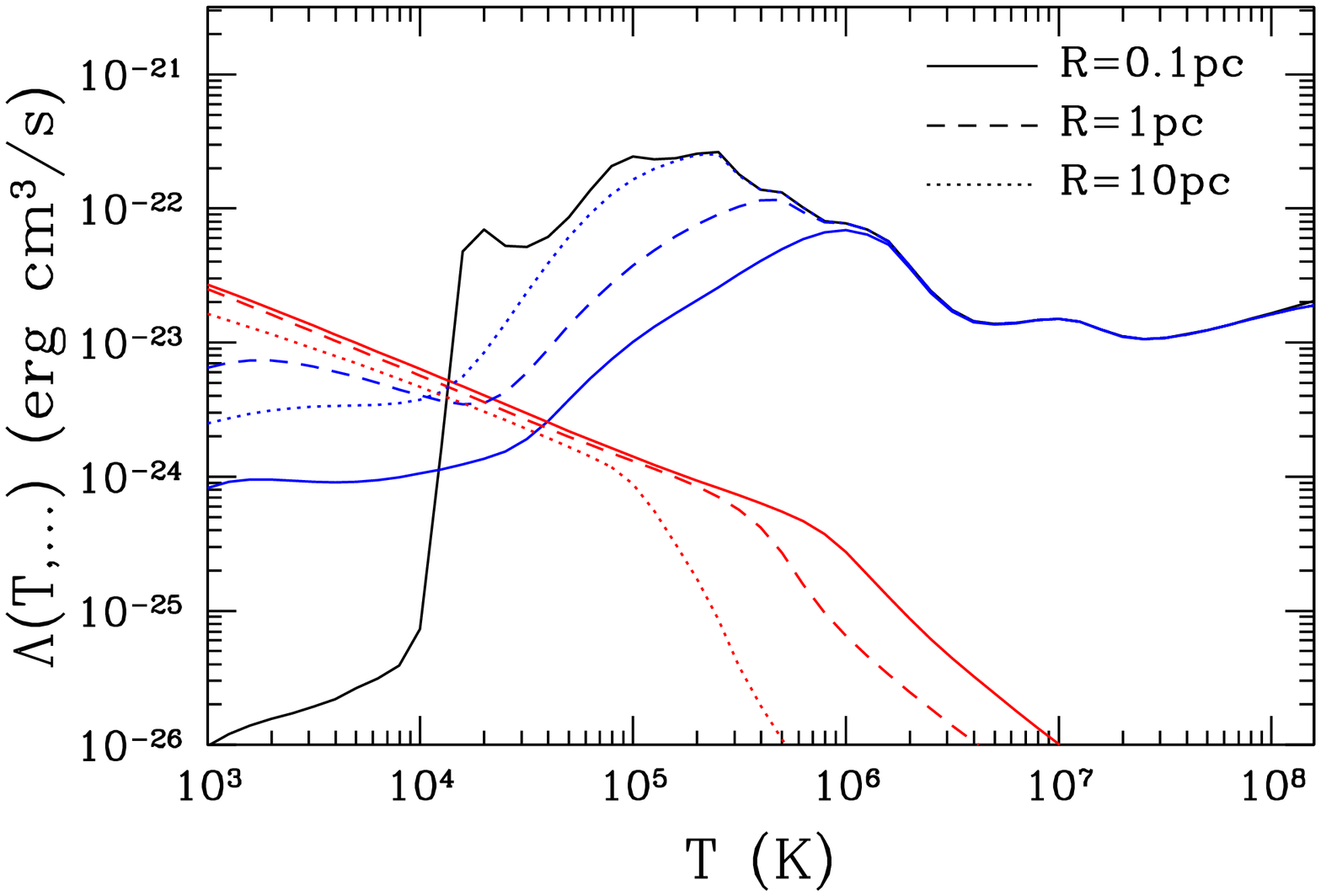}
\caption{\emph{Left}: Cooling (blue lines) and heating (red lines)
functions for a $Z=Z_ \odot$, $n_b = 340\times\bar{n}_b$ galactic halo
at the specified distances from a quasar with an ionizing luminosity
$10^{13}L_\odot$. The black solid line shows the pure CIE ``standard
cooling function''. \emph{Right}: Cooling (blue lines) and heating
(red lines) functions for a $Z=Z_\odot$, $n_b=1\dim{cm}^{-3}$ $\HII$
region around an O star. The black solid line shows the pure CIE
``standard cooling function''.}
\label{fig:qso}
\end{figure}

In the left panel of Figure \ref{fig:qso} we show cooling and heating
functions for a typical galactic halo at $z=0$
($n_b=340\times\bar{n}_b = 8.5\times10^{-5}\dim{cm}^{-3}$) surrounding
a bright quasar with an ionizing luminosity of $10^{13}L_\odot$
(roughly corresponding to a $10^9M_\odot$ black hole). We assume solar
metallicity, a quasar spectrum of $J_\nu \propto \nu^{-2}$, and the
\citet{jnu:hm01} background.

Some interesting consequences may arise from the
radiation-field-dependence of the cooling and heating functions. For
example, gas in the halo within $1\dim{Mpc}$ of this quasar will not
be able to cool and condense into the disk if its virial temperature
is below about $10^5\dim{K}$.

\subsection{$\HII$ Region Around an O Star}
\label{app:ostar}

In the right panel of Figure \ref{fig:qso}, we show the cooling and
heating functions for a solar metallicity cloud with density
$n_b=1\dim{cm}^{-3}$ surrounding an O star with bolometric luminosity
$L=30{,}000L_\odot$. For the stellar spectrum, we assume a black-body
with $T=30{,}000\dim{K}$. The distances we consider are well within
the star's Str\"{o}mgren radius ($\sim30\dim{pc}$), so we may safely
assume that the radial dependence of the starlight is $1/r^2$ (no
depletion due to recombinations). If, instead of a single star, we
consider a cluster of $N$ O stars, our result will still hold if we
simply rescale the distance axis by $N^{1/2}$.

Close enough to the star, the equilibrium temperature of the $\HII$
region can be substantially higher than the canonical $10^4\dim{K}$.

\subsection{Quasar Irradiating its own Halo}
\label{app:agn}

\begin{figure}[h]
\plottwo{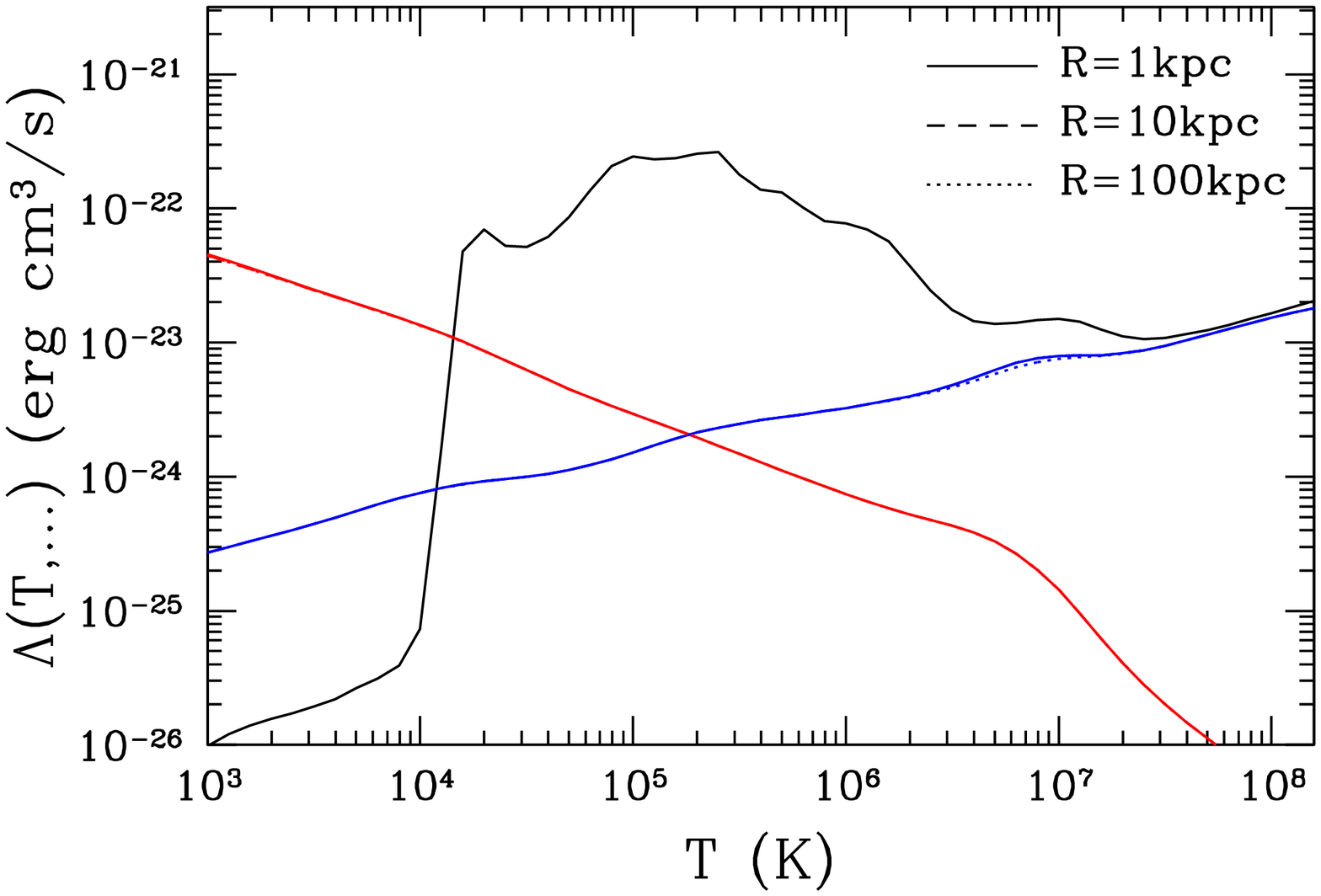}{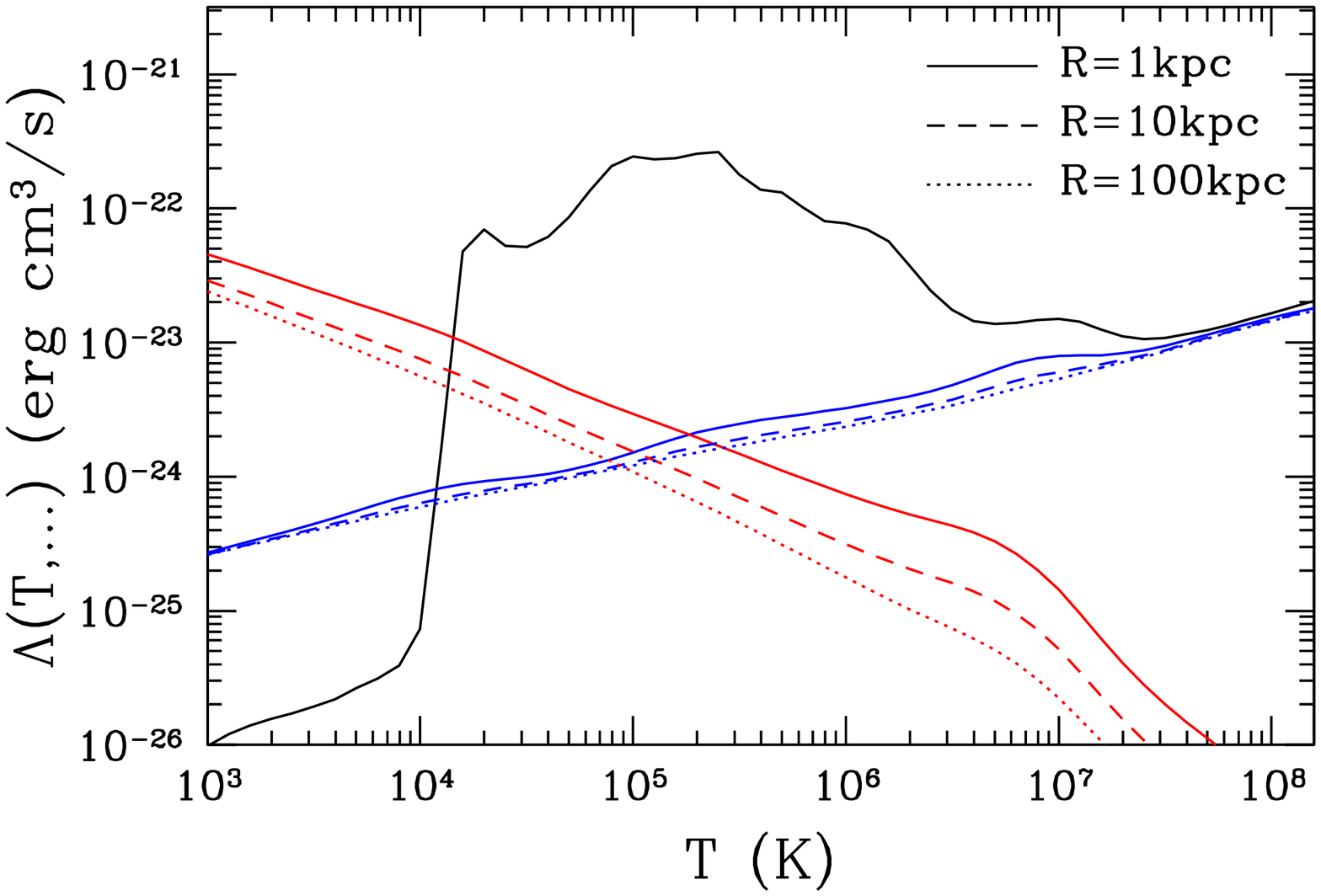}
\caption{Cooling (blue lines) and heating (red lines) curves for a
halo with a $z=3$ isothermal density profile surrounding a quasar with
$10^{13}L_\odot$ in ionizing radiation. In the left panel, the
metallicity is fixed at $0.5Z_\odot$, in which case the cooling and
heating functions become distance-independent. In the right panel, the
metallicity has a mild outward gradient, $Z\propto r^{-1/2}$. The
black solid curves show the pure CIE ``standard cooling function''.}
\label{fig:agn}
\end{figure}

In Figure \ref{fig:agn} we show cooling and heating functions in a
gaseous halo at $z=3$ (virial density $n_b=200\times\bar{n}_b =
3.2\times10^{-3}\dim{cm}^{-3}$) irradiated by a $\sim10^9M_\odot$
central black hole ($10^{13}L_\odot$ in ionizing radiation). The
density profile of the cloud is taken as
\[
  n_b = 3.2\times10^{-3}\dim{cm}^{-3}\left(\frac{100\dim{kpc}}{r}\right)^2
\]
and the metallicity is taken either to be constant $0.5Z_\odot$
(leading to distance-independent heating and cooling functions) or to
have a mild outward gradient,
\[
  Z = 0.5Z_\odot \left(\frac{1\dim{kpc}}{r}\right)^{1/2}.
\]

In both cases, the quasar is capable of maintaining the heating rate
in excess of the cooling rate for $T\la 10^5\dim{K}$. It is therefore
possible to prevent cooling in the halo -- and hence, accretion of
fresh gas onto the galactic disk and the black hole -- without any
need for a mechanical or non-radiative thermal feedback
mechanism. This result is, of course, not new
\citep{atom:socs05,atom:co07}, here we simply use it as an
illustration to the importance of properly accounting for the
radiation field dependence of the cooling and heating functions.

\bibliographystyle{apj}
\bibliography{ng-bibs/atom,ng-bibs/jnu,ng-bibs/self,ng-bibs/cf,ng-bibs/misc,ng-bibs/sims,ng-bibs/ism}

\end{document}